\documentclass[conference]{IEEEtran}
\usepackage[square, numbers, sort&compress]{natbib}
\usepackage{notoccite}

\usepackage{cite}

\usepackage{amsmath}
\usepackage[hidelinks]{hyperref}
\usepackage{dirtytalk}
\usepackage{graphicx}
\usepackage{subfig}
\usepackage{multirow}
\usepackage{lipsum}
\usepackage{booktabs}
\usepackage{longtable}
\usepackage{caption}
\usepackage{adjustbox}
\usepackage{tabularx}
\usepackage{floatrow} 
\usepackage[vlines]{tabularht}
\usepackage{pbox}
\usepackage{pifont}%
\usepackage{multirow}
\usepackage[table,xcdraw]{xcolor}
\usepackage{floatrow}
\ifCLASSINFOpdf
\else
\fi
\begin{document}
%
\title{Grasp-and-Lift Detection from EEG Signal Using Convolutional Neural Network}

\author{\IEEEauthorblockN{Md. Kamrul Hasan$^{\dag, *}$, Sifat Redwan Wahid$^{\dag}$, Faria Rahman$^{\ddag}$, Shanjida Khan Maliha$^{\ddag}$, Sauda Binte Rahman$^{\gamma}$}
\IEEEauthorblockA{$^{\dag}$Department of Electrical and Electronic Engineering (EEE),\\
$^{\ddag}$Institute of Information and Communication Technology (IICT),\\
$^{\gamma}$Department of Biomedical Engineering (BME), \\
Khulna University of Engineering \& Technology (KUET), Khulna-9203, Bangladesh.\\
$^*$Email: m.k.hasan@eee.kuet.ac.bd}
}


%


\maketitle

\begin{abstract}
People undergoing neuromuscular dysfunctions and amputated limbs require automatic prosthetic appliances. In developing such prostheses, the precise detection of brain motor actions is imperative for the Grasp-and-Lift (GAL) tasks. Because of the low-cost and non-invasive essence of Electroencephalography (EEG), it is widely preferred for detecting motor actions during the controls of prosthetic tools. This article has automated the hand movement activity viz GAL detection method from the 32-channel EEG signals. The proposed pipeline essentially combines preprocessing and end-to-end detection steps, eliminating the requirement of hand-crafted feature engineering. Preprocessing action consists of raw signal denoising, using either Discrete Wavelet Transform (DWT) or highpass or bandpass filtering and data standardization. The detection step consists of Convolutional Neural Network (CNN)- or Long Short Term Memory (LSTM)-based model. All the investigations utilize the publicly available WAY-EEG-GAL dataset, having six different GAL events. The best experiment reveals that the proposed framework achieves an average area under the ROC curve of $0.944$, employing the DWT-based denoising filter, data standardization, and CNN-based detection model. The obtained outcome designates an excellent achievement of the introduced method in detecting GAL events from the EEG signals, turning it applicable to prosthetic appliances, brain-computer interfaces, robotic arms, etc.  
\end{abstract}


\IEEEpeerreviewmaketitle

\setlength{\parindent}{2em}

\section{Introduction}
\label{Introduction}
Around 2.1 million people in the United States (US) alone are living with limb injury, and a further 185,000 people demand an amputation each year \citep{cordella2016literature}. Additionally, approximately 300,000 people in the US live with spinal cord injury affecting uppermost extremity function \citep{white2016spinal}. 
According to World Health Organization, about one billion individuals are disabled, with up to 190 million individuals, equating to roughly $15\,\%$ of the world's population. Earlier, persons with impairments could only communicate with a prosthetic voice receiver by speaking, and the prosthesis recognized the signal through the receiver, performing the user's desired action. The patient can also apply an Electromyographic (EMG) signal to accomplish the same goal \citep{sharma2019scalable}. There are several drawbacks to both strategies. The interference from the outside environment can make voice-controlled intelligent prostheses challenging to exercise in public. Besides, several neuromuscular illnesses, such as amyotrophic lateral sclerosis, affect motor neurons, causing the brain to lose control over voluntary muscle movements in the EMG signals \citep{sharma2019scalable}. 
The EMG-based control also contributes to poor dexterity and control versatility, whereas signals from the brain, for example, Electroencephalography (EEG), may produce a more precise alternative and control \citep{gordienko2021deep, hasan2013direct, roy2017eeg}.
However, the decoding of GAL activities from the EEG signal exhibited tremendous recent success in the wrist gestures \citep{li2018decoding}, uppermost limbs \citep{ubeda2017classification}, and elbows \& shoulders \citep{zhou2009eeg}. 
The EEG employs non-invasive electrodes placed on participants' scalps to measure signals produced by local field potentials with active cortex neurons, having high temporal precision. Any detection pipeline, including the decoding of sensation, intention, and action from scalp EEG signal in the WAY-EEG-GAL dataset \citep{luciw2014multi} (see details in Section~\ref{Dataset}), utilize different algorithms, which roughly contain similar essential steps, such as artifact rejection, time-domain filtering, spatial filtering, class feature extraction, and finally, classification \citep{varszegi2016comparison}. 
Several recent works related to hand movement recognition, for instance, GAL events, are reviewed in the subsequent section.

\citet{varszegi2016comparison} contrasted the achievement of three algorithms for GAL event detection. Their first method has four processing stages: artifact rejection, bandpass filter, Common Spatial Pattern (CSP) filter, data normalization. Then, it was inputted to Logistic Regression (LR) ensemble. The second algorithm had three preprocessing steps: artifact elimination, lowpass filtering, and data standardization. The third method utilized artifact rejection and normalization, then inputted to the 1D Convolutional Neural Network (CNN). \citet{wang2006common} proposed scalp mapping-based channel selection and combined it with event-related desynchronization for the detection of imagery hand \& foot movement. In \citep{park2017filter}, the authors introduced subband regularized CSP for GAL classification, using a filter bank for breaking EEG signals in 4–40 Hz range into 4 Hz subbands. \citet{liao2017major} connected both CSP and filter bank for multi-class labeling. They separated the EEG signal into multiple frequency bands using a filter bank for CSP feature extraction.
\citet{singhal2019classification} employed CNN and Butterworth lowpass filter for GAL event detection tasks. 
Their network had five layers, including the data input layer, 1D convolutional layer with pooling layer (max), and finally,  two fully connected layers. However, some of these methods employed extensive hand-crafted feature engineering, which is very difficult to generate a robust model, requiring massive parameter tuning \citep{hasan2020automatic}. 
Although some other strategies alleviate the necessity of feature engineering, there is still room for performance improvement. Keep this in mind; this article has automated the GAL detection task without feature engineering, bestowing better results for the same purpose and dataset.

This paper is designed as follows: Section~\ref{Methodologies} covers the materials and methodology. Section~\ref{Results_and_Discussion} provides the experimental results. Finally, Section~\ref{Conclusion} concludes the article with the future working directions.

\section{Materials and Methodologies}
\label{Methodologies}
Firstly, we explain the utilized dataset in Section~\ref{Dataset}. Secondly, we describe the proposed methods in Section~\ref{ProposedFramework}. 

\subsection{Dataset}
\label{Dataset}
\begin{figure*}[!ht]
  \centering
\includegraphics[width=16cm, height= 8.5cm]{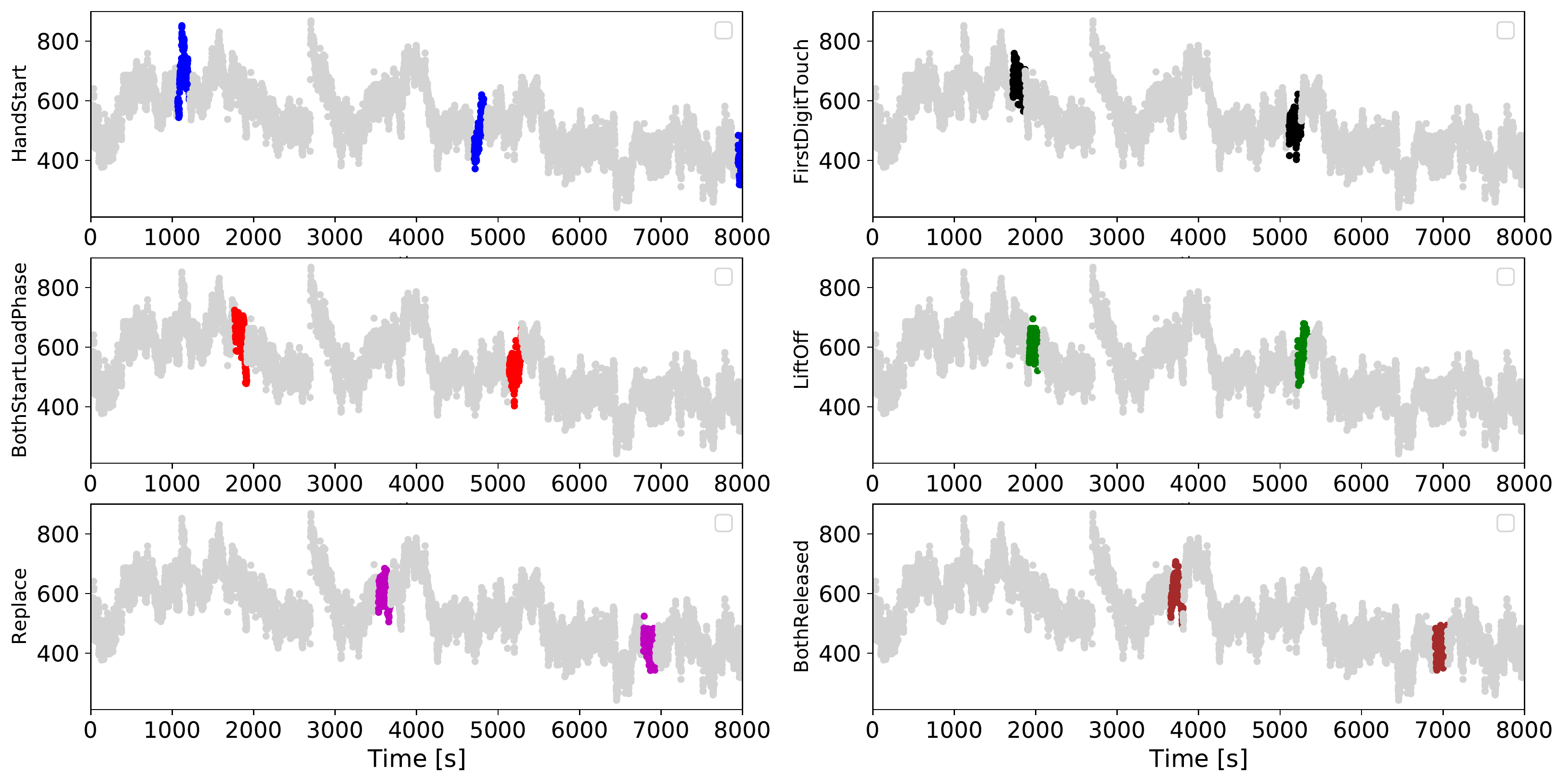}
\caption{Exhibition of all events in the used dataset for subject-1, conferring the O1 channel's EEG signal out of 32 channels.}
\label{fig:all_Events}
\end{figure*}

The utilized WAY-EEG-GAL dataset \citep{luciw2014multi} is intended to decode sensation, intention, and action from scalp EEG signals. Twelve subjects attended the study, where the grasping and lifting object's weight ($165$, $330$, or $660$ gram), surface friction (sandpaper, suede, or silk surface), or both were modified unpredictably (see details in \citep{luciw2014multi}). The WAY-EEG-GAL dataset contains 32-channels EEG signal, having six different events, which are enlisted in Table~\ref{tab:GAL_Events}. 
\begin{table}[!ht]
\caption{Details of six events in the WAY-EEG-GAL dataset. }
\label{tab:GAL_Events}
\centering
\begin{tabular}{p{2.8cm}|p{5.2cm}}
\hline
\rowcolor[HTML]{C0C0C0} 
GAL Events & Description \\ \hline
HandStart (HS)           &   Reaching for the object          \\ \hline
FirstDigitTouch (FDT)          &    Grasp the object using thumb and index finger         \\ \hline
BothStartLoadPhase (BSP)          &   Lifting an object for a couple of seconds          \\ \hline
LiftOff (LO)         &      Lift force       \\ \hline
Replace (R) & Set the object backward on the support surface  \\ \hline
BothRelease (BR) & Free the object and place hand at the starting location \\ \hline
\end{tabular}
\end{table}
Those six different events in the utilized dataset are displayed in Fig.\ref{fig:all_Events}, where each event exist at various time of the O1 channel of the utilized EEG signal.  

\subsection{Proposed Framework}
\label{ProposedFramework}
The proposed automated pipeline essentially consists of a preprocessing (described in Section~\ref{Preprocessing}) and a detection model (described in Section~\ref{Detection}). In preprocessing, we employ different types of filtering, such as Butterworth and Discrete Wavelet Transform (DWT) filters, to remove various artifacts. We also standardize the data to reduce the data skewness \citep{hasan2020diabetes}. On the other hand, we apply two end-to-end networks based on CNN and Long Short Term Memory (LSTM) in the detection model. However, those methods are elaborately explained in the following sections.

\subsubsection{Proposed Preprocessing}
\label{Preprocessing}
Raw EEG signal's ($x$) DWT is determined by transferring it into a filter series. Firstly, the specimens are given through a lowpass filter, having a impulse response of $g$, resulting in a convolution of the two as in (\ref{eq:conv}). 
\begin{equation}
    \label{eq:conv}
    y[n] = (x * g)[n] = \sum\limits_{k =  - \infty }^\infty  {x[k] g[n - k]}
\end{equation}
\noindent Then, the EEG signal $x$ is also decomposed simultaneously applying a highpass filter $h$. In DWT, EEG signal $x$ is filtered with highpass and lowpass filter banks. The lowpass subband is iteratively refined through the lowpass and highpass filters, and its coefficients are selected by the mother wavelet, such as the Haar, Daubechies, Coiflets Symlets, and Bi-orthogonal wavelets. 
In our framework, we incorporate Daubechies 1 (db1) and Daubechies 2 (db2) for ablation studies. Furthermore, Butterworth filters have a maximally flat frequency response feature and no ripples in the passband, rolling towards zero in the stopband. We unite two variants of Butterworth filters, such as Highpass Filter (HPF) and Bandpass Filter (BPF), with the order of $5$, which gives a cut-off frequency of $f_{cut}$ as in (\ref{eq:fcut}).  
\begin{equation}
    \label{eq:fcut}
f_{cut}(x) = \frac{1}{1+(\lambda \times \sigma)^5}
\end{equation}
\noindent where $0<\lambda<1$ is a time steps for controlling the degree of smoothing.  Finally, the standardization ($R$) is fulfilled using $R(x)=\frac{x-\bar{x}}{\sigma}$, where $x$ is the $n$-dimensional instances of the feature vector, $x\in R^n$. $\bar{x}\in R^n$ and $\sigma \in R^n$ is the mean and standard deviation.

\subsubsection{Proposed Detection Model}
\label{Detection}
\begin{figure}[!ht]
  \centering
\includegraphics[width=8.5cm, height= 2.9cm]{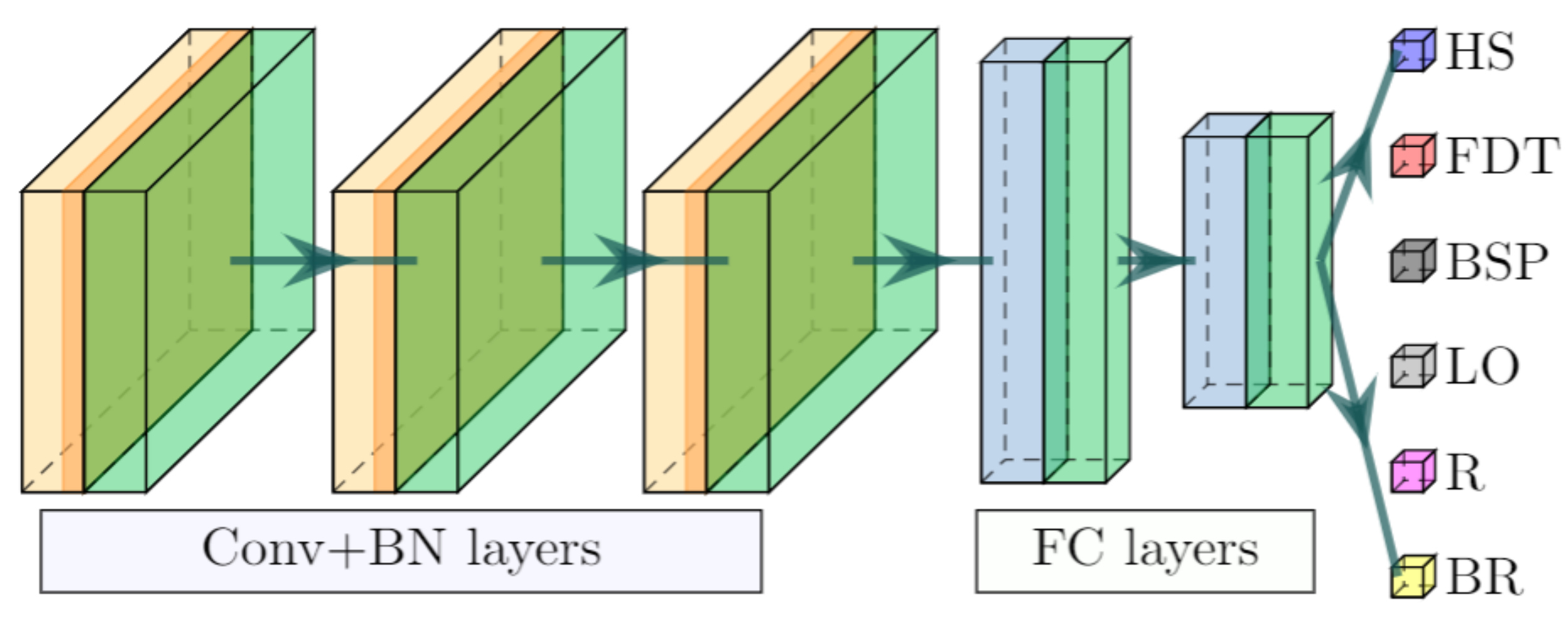}
\caption{The recommended CNN-based end-to-end GAL event detection model, which takes an EEG signal as an input and produces the score of each event as an output. }
\label{fig:CNN}
\end{figure}

\begin{figure}[!ht]
  \centering
\includegraphics[width=8.5cm, height= 2.5cm]{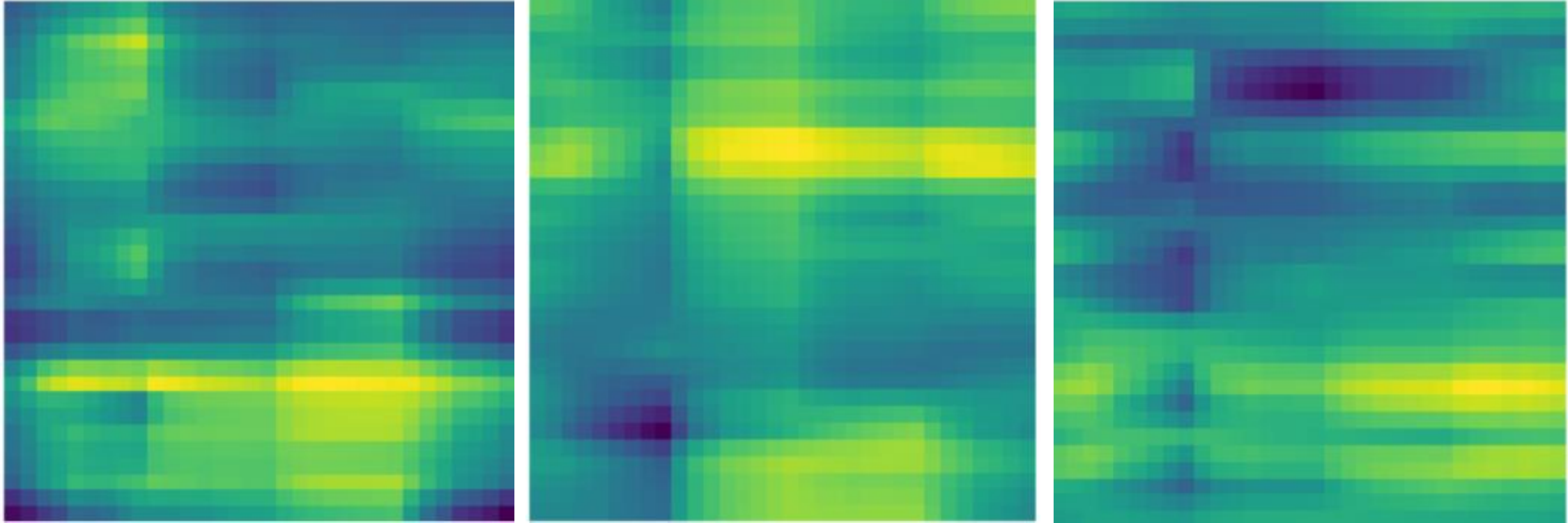}
\caption{Example of generated 2D samples from the 1D EEG signal of the WAY-EEG-GAL dataset.}
\label{fig:CNN_samples}
\end{figure}

\begin{figure}[!ht]
  \centering
\includegraphics[width=7cm, height= 4cm]{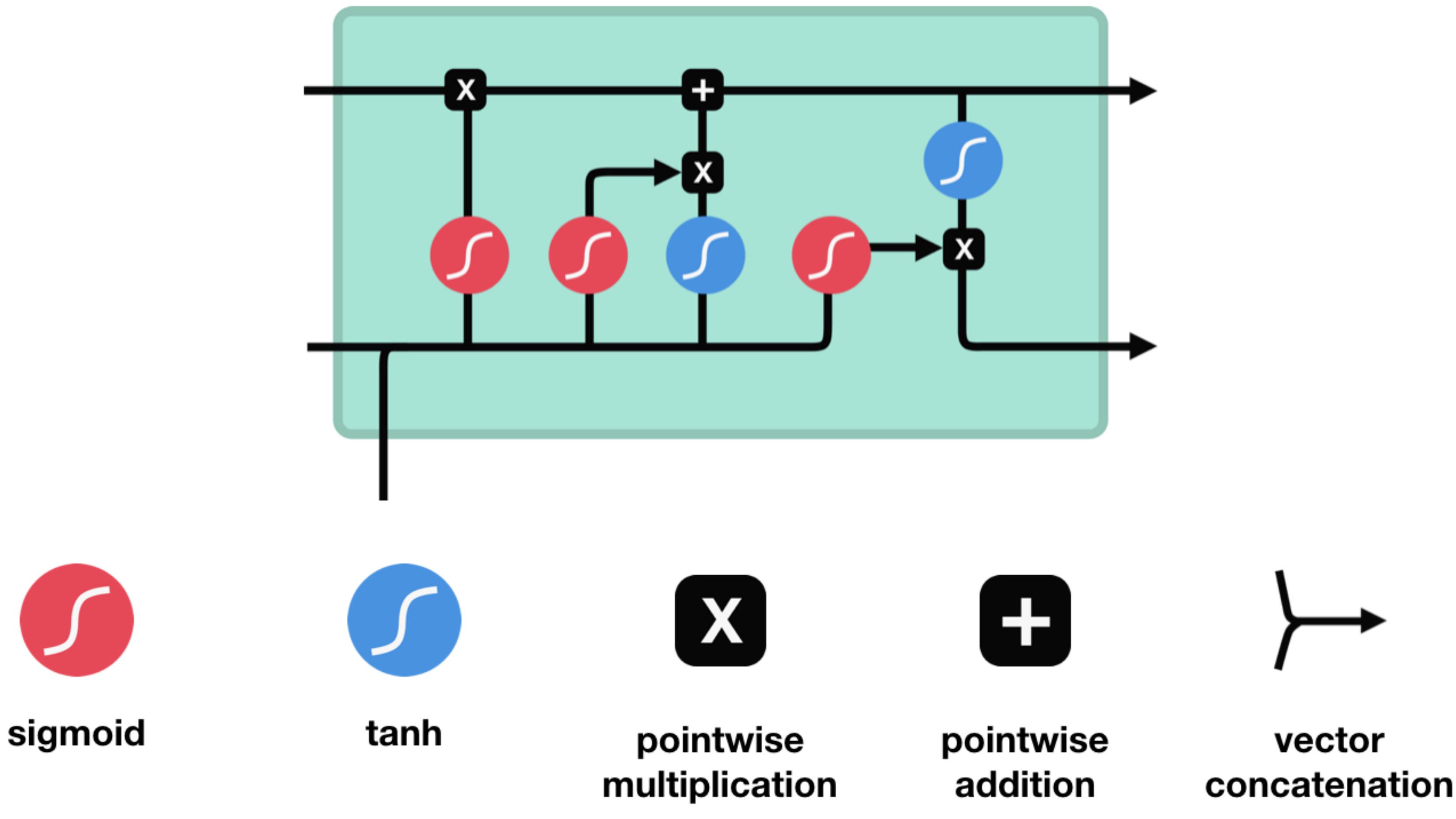}
\caption{The employed LSTM cell \citep{wang2018lstm} in the proposed LSTM-based end-to-end GAL event detection model.}
\label{fig:LSTM}
\end{figure}

The denoised EEG signals are then processed for automatic GAL event detection. We have employed two detection models, CNN and LSTM based, for comparative analysis. The former CNNs use filters within convolutional layers to transform input data, which is better for spatial information processing. In contrast, the latter LSTMs reuse activation functions from other data points in the sequence to generate the following output in a series, which is more suitable for temporal information processing. Fig.~\ref{fig:CNN} shows the proposed CNN-based detection model.
The combination of Convolution (Conv), with ReLU activation, and Batch Normalization (BN) layers (Conv+BN) learn discriminating features \citep{hasan2021detection} from the inputted EEG signal about the GAL events. The BN layers are joined with the Conv layers to minimize the internal covariate shift during each mini-batch iteration. Then, the Fully-connected layers (FC) classifies those learned features into GAL event categories (see in Fig.~\ref{fig:CNN}). This CNN network is trained with the 2D samples generated from the 1D EEG signals (see examples in Fig.~\ref{fig:CNN_samples}). 
On the other hand, the proposed LSTM-based model is composed of LSTM cells (see in Fig.~\ref{fig:LSTM}) and dropout layers, which are repeated four times in a cascaded fashion.
Finally, the FC layer with six neurons classifies the target GAL events. 

\subsubsection{Training and Evaluation}
\label{Training_and_Evaluation}
The experiments are carried out in a \textit{Windows-10} machine, using different Python and Keras APIs. The utilized computer has the following hardware configuration:
Intel\textsuperscript{\tiny\textregistered} Core\textsuperscript{\tiny{TM}} i$7$-$7700$ HQ CPU @ $2.80\,GHz$ processor with Install memory (RAM): $16.0\,GB$ and GeForce GTX $1060$ GPU with $6\,GB$ GDDR$5$ memory. 
The obtained results are investigated using the Receiver Operating Characteristic (ROC) curve and Area Under the ROC Curve (AUC) value, which dispenses the performance of a classification model at various thresholds.

\section{Results and Discussion}
\label{Results_and_Discussion}
This section is dedicated to presenting the obtained results from different extensive experiments. Firstly, we demonstrate our experimental results in the manner of ablation studies. Secondly, we have compared our proposed framework with several published articles for the same task and dataset. 

\begin{figure*}[!ht]
  \centering
\includegraphics[width=16cm, height= 5.9cm]{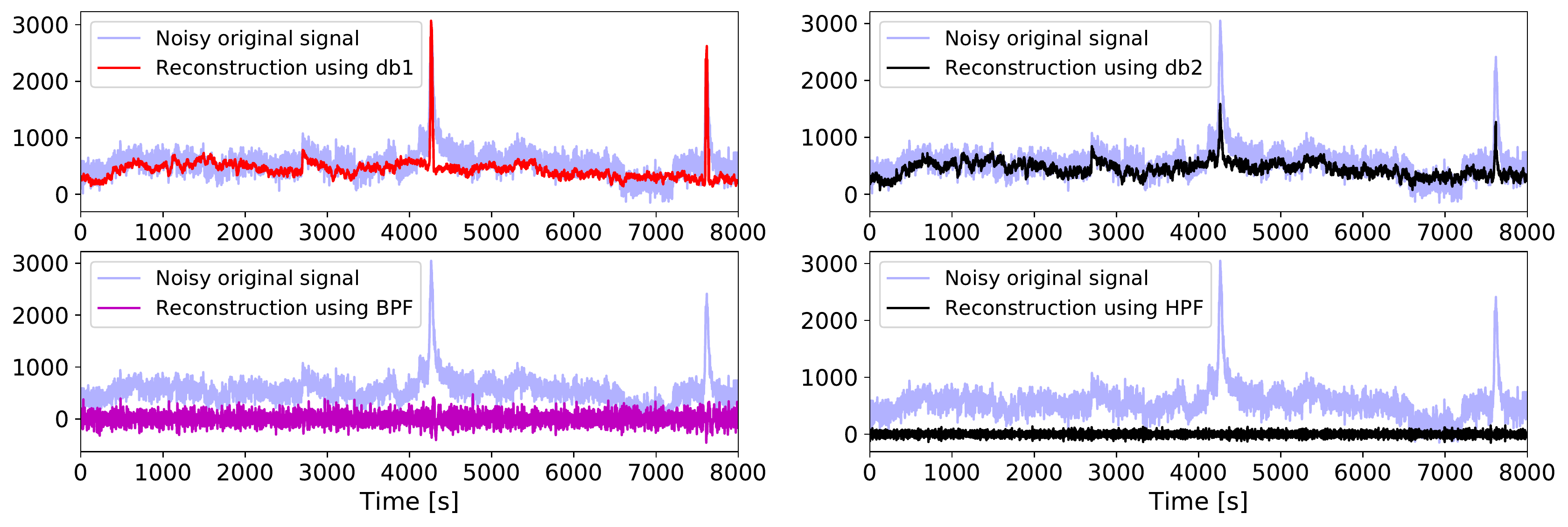}
\caption{Denoising of raw EEG signal for further detection stage in the proposed framework, employing two variants of DWT filters (top-left and right) and Butterworth BPF and HPF filters (bottom-left and right).}
\label{fig:Noise_reduction}
\end{figure*}

The raw EEG signal suffers from various noises originating from the muscle, eye movement and blinking, power lines, and other devices' interference. 
Therefore, they need to be removed essentially, resulting in an accurate estimation.
The obtained results from four different methods in the proposed framework are exhibited in Fig.~\ref{fig:Noise_reduction}. 
The denoise results in Fig.~\ref{fig:Noise_reduction} qualitatively reveal that the DWT filters preserve the signal properties comparing the Butterworth filters (HPF and BPF). There is little-to-no signal leakage or phase-shifting of the original signal in wavelet-based filtering. In contrast, the Butterworth filters have problems with signal leakage, phase-shifting, and fixed-coefficient. Therefore, the former DWT filters outperform the conventional Butterworth filters (see in Fig.~\ref{fig:Noise_reduction}). Furthermore, comparing db1 and db2 mother wavelets in DWT filters (top row in Fig.~\ref{fig:Noise_reduction}) qualitatively, it can be concluded that db2 reduces the sudden peak vastly by retaining signal shape information. Therefore, in the upcoming experiments for automatic GAL event detection, we consider only DWT filters with db2 mother wavelet for denoising the raw EEG signal.

\begin{figure*}[!ht]
  \centering
\subfloat[Without preprocessing on CNN-based model]{\includegraphics[width=8.5cm, height= 6cm]{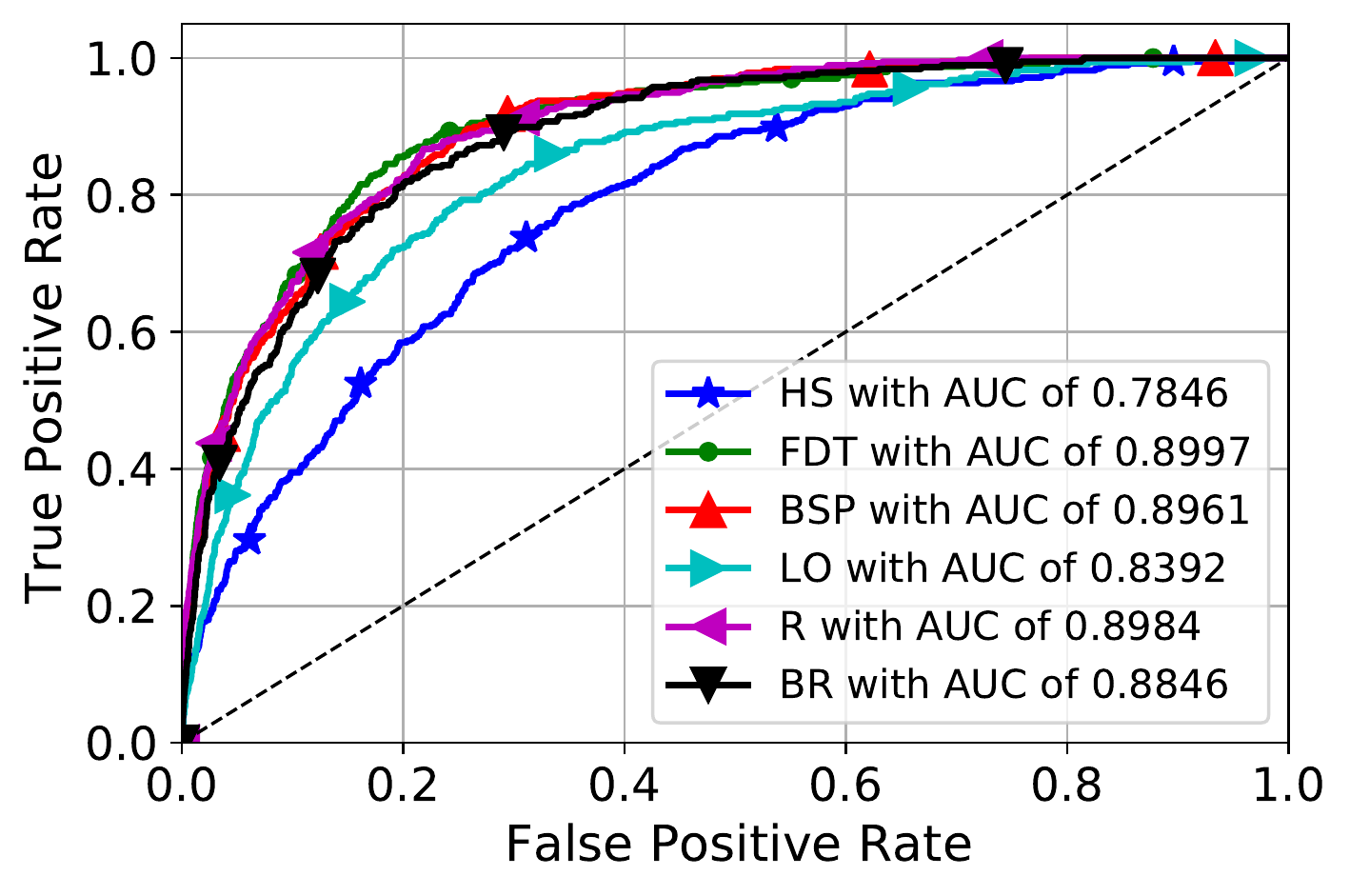}}
\subfloat[With preprocessing on CNN-based model]{\includegraphics[width=8.5cm, height= 6cm]{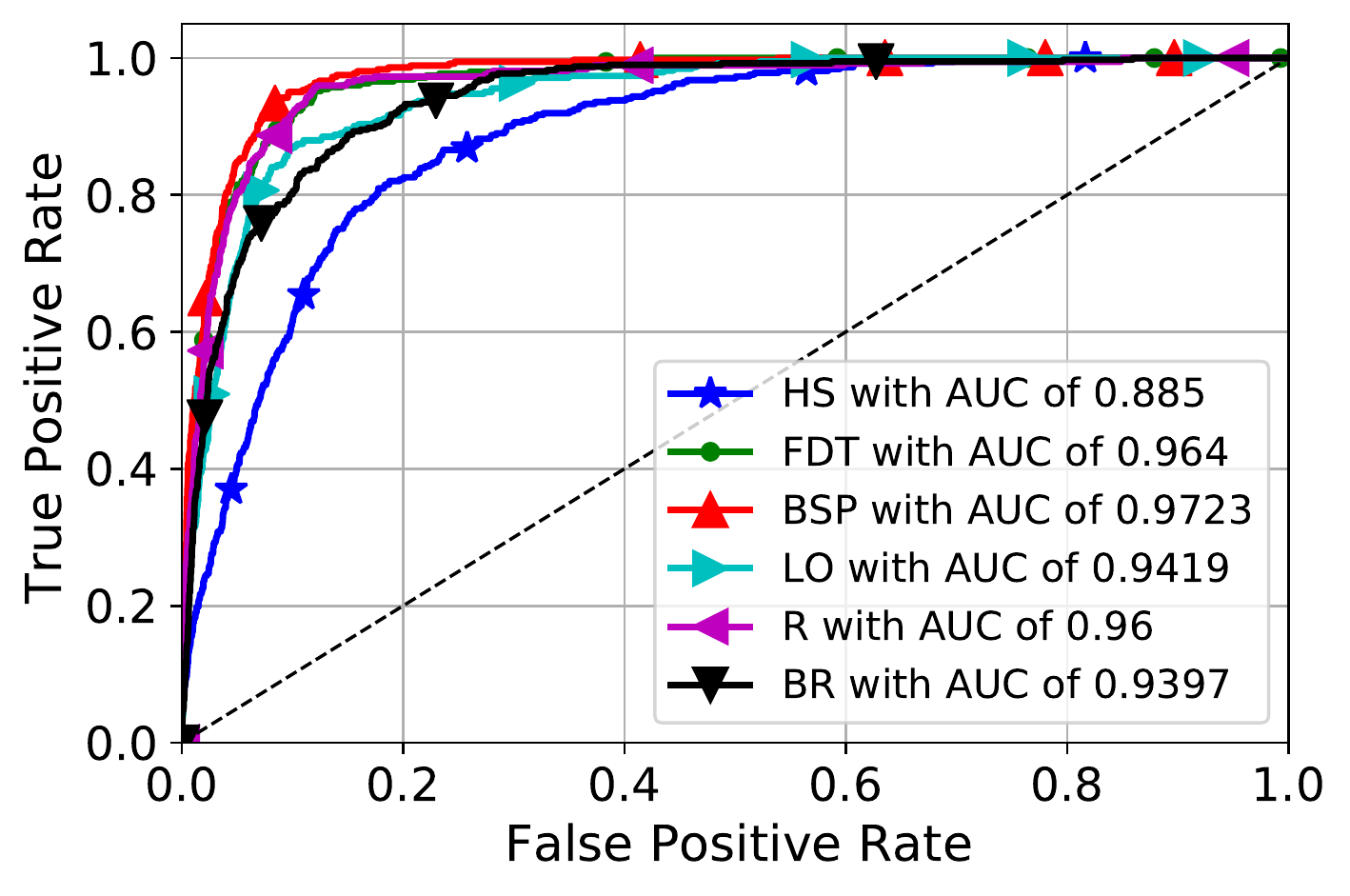}} \\
\subfloat[Without preprocessing on LSTM-based model]{\includegraphics[width=8.5cm, height= 6cm]{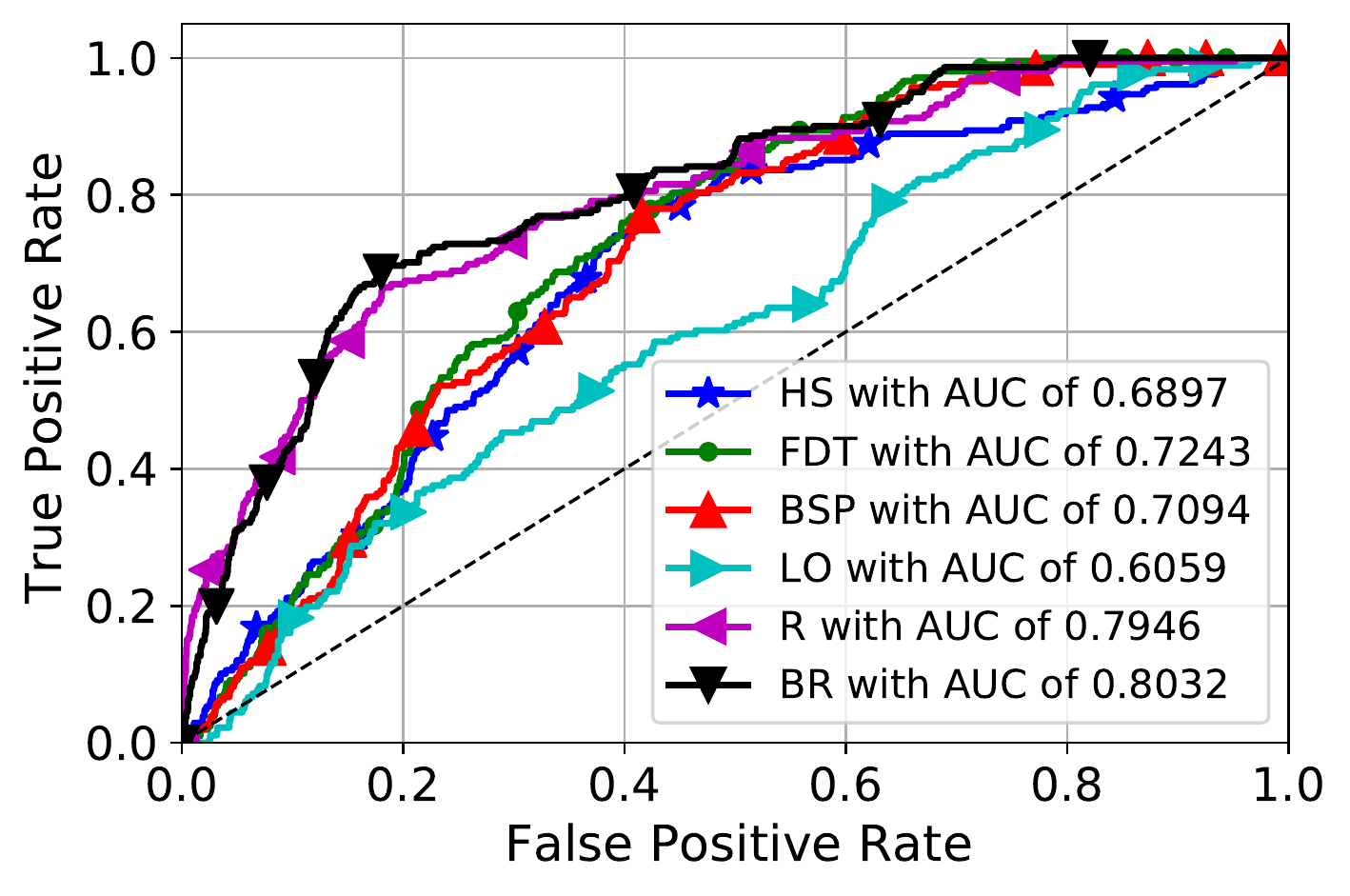}}
\subfloat[With preprocessing on LSTM-based model]{\includegraphics[width=8.5cm, height= 6cm]{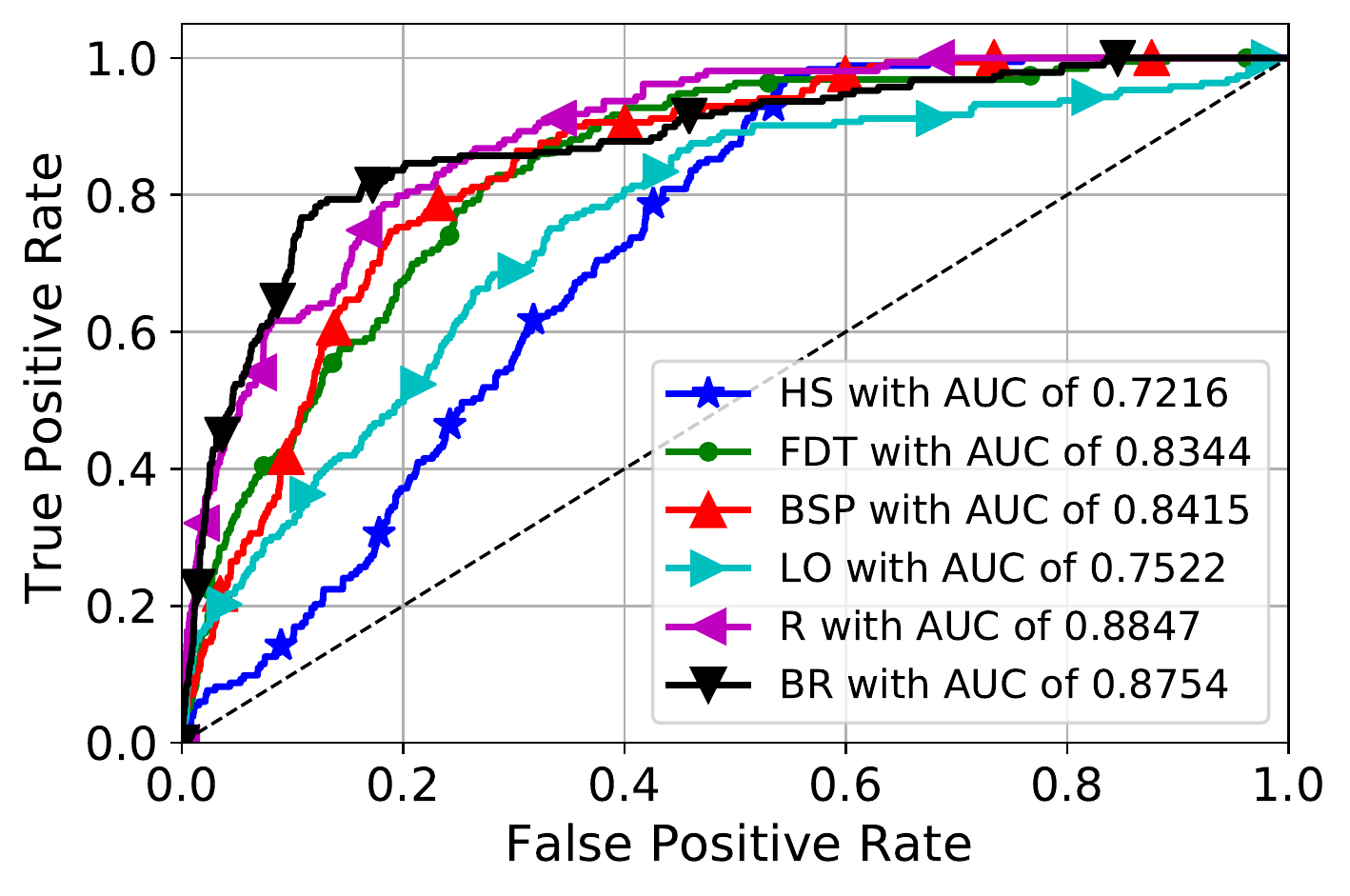}}
\caption{ROC curves for GAL events recognition with the corresponding AUC values from the CNN-based (top row) and LSTM-based (bottom row) models, with (right column) and without (left column) proposed integrated preprocessing.}
\label{fig:ROC_CNN_LSTM}
\end{figure*}

The EEG signals after denoising are directly fed to the detection model, either CNN- or LSTM-based, for automatic event prediction.
Table~\ref{tab:Detection_Results} displays the quantitative results for different experiments, where we demonstrated the results for with/without the proposed preprocessing for establishing the efficacy of the recommended preprocessing. 
\begin{table*}[!ht]
\centering
\caption{GAL event detection results using CNN- and LSTM-based models, employing a proposed preprocessing method, where we publicize AUC as an evaluation metric. The best-obtained values are in bold fonts.}
\label{tab:Detection_Results}
\begin{tabular}{ccccccccc}
\hline
\rowcolor[HTML]{C0C0C0} 
\cellcolor[HTML]{C0C0C0}                                                                   & \cellcolor[HTML]{C0C0C0}                         & \multicolumn{6}{c}{\cellcolor[HTML]{C0C0C0}AUC of different events} & \cellcolor[HTML]{C0C0C0}                                                                        \\ \cline{3-8}
\rowcolor[HTML]{C0C0C0} 
\multirow{-2}{*}{\cellcolor[HTML]{C0C0C0}Processing}                                       & \multirow{-2}{*}{\cellcolor[HTML]{C0C0C0}Models} & HS        & FDT        & BSP        & LO       & R       & BR       & \multirow{-2}{*}{\cellcolor[HTML]{C0C0C0}\begin{tabular}[c]{@{}c@{}}Average\\ AUC\end{tabular}} \\ \hline
                                                                                           & CNN                                             &    $0.785$      &  $0.900$         &   $0.896$       &  $0.839$       &   $0.898$     &   $0.885$      & $0.867$                                                                                                  \\ \cline{2-9}
\multirow{-2}{*}{\begin{tabular}[c]{@{}c@{}}Without proposed\\ preprocessing\end{tabular}} & LSTM                                             &  $0.690$        &      $0.724$      &    $0.709$        &    $0.606$      &   $0.795$      &   $0.803$       &     $0.721$                                                                                             \\ \hline
                                                                                           & CNN                                            &    $\mathbf{0.885}$      &  $\mathbf{0.964}$         &   $\mathbf{0.972}$       &  $\mathbf{0.942}$       &   $\mathbf{0.960}$     &   $\mathbf{0.940}$      & $\mathbf{0.944}$                                                                                                \\  \cline{2-9}
\multirow{-2}{*}{\begin{tabular}[c]{@{}c@{}}With proposed\\ preprocessing\end{tabular}}    & LSTM                                             &   $0.722$        &        $0.834$    &    $0.842$        &   $0.752$       &    $0.885$     &   $0.875$       &    $0.818$                                                                                             \\ \hline
\end{tabular}
\end{table*}
The experimental results show that the proposed preprocessing increases the average AUC values with the margins of $7.7\,\%$ and $9.7\,\%$, respectively, for the CNN- and LSTM-based models.
Again, the results of HS, FDT, BSP, LO, R, and BR events have been boosted by $10.0\,\%$, $6.4\,\%$, $7.6\,\%$, $10.3\,\%$, $6.2\,\%$ and $5.5\,\%$, respectively, for the CNN-based model due to the appliance of proposed preprocessing. Those values are $3.2\,\%$, $11.0\,\%$, $13.3\,\%$, $14.6\,\%$, $9.0\,\%$, and $7.2\,\%$ for LSTM-based model. 
Significantly, the AUCs of LO event for both the detection models have been enhanced by the border of $10.3\,\%$ and $14.6\,\%$. 
Furthermore, the same patterns of AUC enhancement due to the integration of the recommended preprocessing are noticed in the ROC curves in Fig.~\ref{fig:ROC_CNN_LSTM}. Those figures illustrate that both the detection models improve the ROC curves and their AUC values when raw EEG signals are processed to remove different noises, such as muscle, eye movement and blinking, power line, and interference with other devices. Therefore, those discussions on the achieved results for the GAL event detection appropriately prove the supremacy of the proposed preprocessing for the intended task. The experiments also confirm that the different artifacts in the extracted EEG signal adversely affect the GAL event detection results. 

\begin{figure}[!ht]
  \centering
\includegraphics[width=8.5cm, height= 5.5cm]{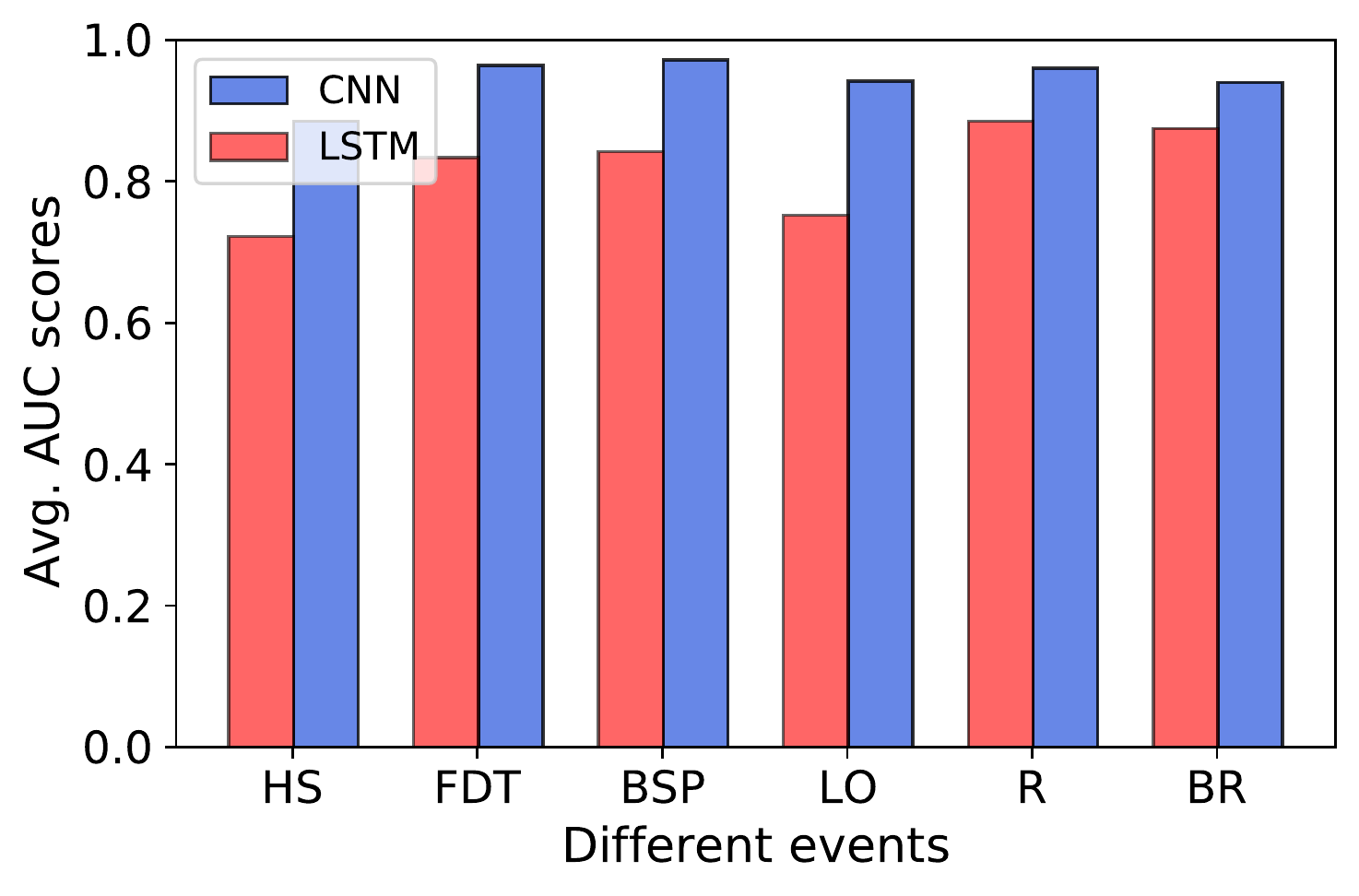}
\caption{Quantitative comparison of the achieved results in terms of AUC values from the CNN- and LSTM-based event detection model, employing the same proposed pipeline.}
\label{fig:ROC_compare}
\end{figure}

Again, further investigation on the CNN and LSTM results shows that the CNN-based model betters the LSTM-based model (see in Table~\ref{tab:Detection_Results} and Fig.~\ref{fig:ROC_CNN_LSTM}). The results of the CNN-based model in Fig.~\ref{fig:ROC_CNN_LSTM} point that it outperforms the LSTM model with the margins of $16.3\,\%$, $13.0\,\%$, $13.0\,\%$, $19.0\,\%$, $7.5\,\%$, and $6.5\,\%$ respectively for the HS, FDT, BSP, LO, R, and BR events in terms of AUC value (see in Fig.~\ref{fig:ROC_compare}). 
The ROC curves in Fig.~\ref{fig:ROC_CNN_LSTM} of the CNN- and LSTM-based models also reveal the more excellent performance of the former model, dispensing better AUCs from the corresponding curves. 
Due to the chain rule application in LSTM training, while calculating the error gradients, the domination of the multiplicative term increases over time, which forces the LSTM model's gradient to be exploded or vanished. Such a limitation of the LSTM model is the possible reason for getting less performance for the GAL detection in our experiments. Furthermore, the CNN has feature parameter sharing and dimensionality reduction, which reduces the number of parameters; thus, the computations are also decreased in the CNN model comparing the LSTM model. Those discussions reveal the power of the recommended CNN-based model for the GAL event detection over the LSTM model. 

Many articles \citep{varszegi2016comparison, sharma2019scalable, yahya2019classification, gordienko2021deep} have already been published for the same task on the WAY-EEG-GAL dataset. Comparing them in terms of AUC, our proposed pipeline has outputted better results than the works of \citet{sharma2019scalable}, \citet{gordienko2021deep}, and \citet{varszegi2016comparison} with the margins of $9.4\,\%$, $2.4\,\%$, and $11.5\,\%$, respectively. Besides, the proposed detection system alleviates the necessity of hand-crafted feature extraction and engineering like \citep{roy2017eeg, sharma2019scalable, varszegi2016comparison}. Although the outcome of \citet{yahya2019classification} provides a better value of AUC, they used extensive preprocessing and deeper network architecture with roughly $6.8$ million parameters. In contrast, our proposed network has only $1.4$ million, making it more applicable in real-time applications in prosthetic appliances, brain-computer interfaces, robotic arms, etc.

\section{Conclusion and Future Works}
\label{Conclusion}
This article has automated the GAL event detection task for the publicly available WAY-EEG-GAL dataset. The recommended preprocessing has enhanced the outcome by a significant margin, especially $7.7\,\%$ for the CNN model. The DWT-based filtering with Daubechies 2 mother wavelet has retained the signal property by reducing sudden peak noise is the possible reason for such an enhancement. Again, the proposed CNN-based GAL detection model has outputted better results than the other LSTM-based model, notably $12.6\,\%$, as the former has satisfactory feature learning capability. Furthermore, our experimental results have pointed out that the CNN can produce what LSTM has been applied for and is excellent at predicting the events in the WAY-EEG-GAL dataset but in a much faster, more computationally effective fashion. The layers, kernel size with its initialization, and others of the proposed CNN-based detection model will be tuned in the future for getting the best network for the same task and dataset to enhance the outcomes. The optimum channels from the 32-channel EEG signal will also be estimated to reduce the computational burden for the real-time appliances. In the future, the suggested framework will also be employed in the other detection problems to validate its efficacy and versatility.

\bibliographystyle{model2-names}

\bibliography{sample}

\end{document}